\documentclass[showpacs,showkeys,eprintnumbers,amsmath,amssymb,
  nofootinbib]{revtex4}
\usepackage{amsmath}
\usepackage{amssymb}
\usepackage{amsfonts}
\usepackage{eucal}
\usepackage{graphicx}
\newcommand{\be}{\begin{equation}}
\newcommand{\bse}{\begin{subequations}}
\newcommand{\ese}{\end{subequations}}
\newcommand{\ba}{\begin{eqnarray}}
\newcommand{\ea}{\end{eqnarray}}
\newcommand{\ee}{\end{equation}}

\begin{document}
\begin{flushright}
IPM/P-2007/005
\end{flushright}
\begin{center}

\end{center}
\vspace*{2mm}

\title{Tracing CP-violation in Lepton Flavor Violating Muon Decays }

\author{Yasaman Farzan}
\email{yasaman@theory.ipm.ac.ir}

\affiliation{  Institute for Studies in Theoretical Physics and
  Mathematics (IPM)
P.O. Box 19395-5531, Tehran, Iran\\}

\begin{abstract}
Although the Lepton Flavor Violating (LFV) decay $\mu^+\to e^+
\gamma$ is forbidden in the Standard Model (SM), it can take place
within various  theories beyond the SM. If the branching ratio of
this decay saturates its  present bound  [{\it i.e.,} Br$(\mu^+
\to e^+\gamma)\sim 10^{-11}$], the forthcoming experiments can
measure the branching ratio with high precision and consequently
yield information on the sources of LFV. In this letter, we show
that for polarized $\mu^+$, by studying the angular distribution
of the transversely polarized positron and linearly polarized
photon we can derive information on the CP-violating sources
beyond those in the SM. We also study the angular distribution of
the final particles in the decay $\mu^+\to e^+_1 e^- e^+_2$ where
$e^+_1$ is defined to be the more energetic positron.  We show
that transversely polarized $e_1^+$ can provide information on a
certain combination of the CP-violating phases of the underlying
theory which would be lost by averaging over the spin of $e^+_1$.
\end{abstract}

\date{\today}
\maketitle

\section{Introduction}
While the Standard Model (SM) preserves the lepton flavor,  its
various extensions such as supersymmetry or large extra dimensions
can lead to Lepton Flavor Violating (LFV) rare decays $\mu^+ \to
e^+ \gamma$ and $ \mu^+ \to e^+ e^-e^+$ detectable in the
forthcoming experiments. The present experimental bounds on the
branching ratios of these processes are \cite{pdg}
$${\rm Br}(\mu^+ \to e^+ \gamma)<1.2 \times 10^{-11} \ \ \ \ {\rm
Br}(\mu^+ \to e^+ e^+e^-)<1.0\times 10^{-12} \ \ {\rm at} \  90\%
\ {\rm C.L.}$$
 The MEG experiment at PSI \cite{MEGhomepage}, which
is under construction, will be able to probe Br$(\mu^+ \to e^+
\gamma)$ down to $10^{-14}$. Thus, if this branching ratio
saturates the present bound ({\it i.e.,} Br$(\mu^+ \to e^+
\gamma)\sim 10^{-11}$) the future searches will enjoy high
statistics and can make precise measurement limited only by
systematics. Moreover, since the muons are produced by decay of
stopped pions (at rest), they will be almost 100\% polarized.
Thus, studying the angular distribution of the final positrons, we
can learn about phenomena such as parity violation, through which
more information on the sources of LFV can be
extracted~\cite{review}.  Among the various extensions of the SM
that can give rise to lepton flavor violating phenomena, in the
literature the Minimal Supersymmetric Standard Model (MSSM) and
large extra dimensions have received particular attention.  It is
well-known that in both cases integrating out the heavy states of
the model, the LFV Lagrangian responsible for $\mu \to e \gamma$
can be written as \be {\cal L}=A_R \bar{\mu}_R \sigma^{\mu \nu}e_L
F_{\mu \nu}+A_L \bar{\mu}_L \sigma^{\mu \nu} e_R F_{\mu \nu}+A_R^*
\bar{e}_L \sigma^{\mu \nu}\mu_R F_{\mu \nu}+A_L^* \bar{e}_R
\sigma^{\mu \nu} \mu_L F_{\mu \nu} \label{LFV-lagrangian}\ee where
$\sigma^{\mu \nu}=\frac{i}{2}[\gamma^\mu,\gamma^\nu]$ and $F_{\mu
\nu}$ is the photon field strength: $F_{\mu \nu}=\partial_\mu
\varepsilon_\nu -\partial_\nu \varepsilon_\mu$. Although
Lagrangian in Eq.~(\ref{LFV-lagrangian}) is not the most general
form of the effective LFV Lagrangian, throughout this paper we
consider only these terms. In the appendix, we consider  a more
general form of the effective Lagrangian and show that
(\ref{LFV-lagrangian}) is indeed the dominant part.


 It can be shown that averaging over the spins of the
final particles, Lagrangian~(\ref{LFV-lagrangian}) yields
\cite{veto}\be \frac{d\Gamma(\mu^+ \to e^+ \gamma)}{d\cos
\theta}=\frac{1}{8 \pi} m_\mu^3\left[|A_R|^2(1-\mathbb{P}_\mu \cos
\theta)+|A_L|^2( 1+\mathbb{P}_\mu \cos
\theta)\right],\label{averaging-spin}\ee where $\theta$ is the
angle between the momentum of the positron and the spin of the
muon and $\mathbb{P}_\mu$ is the polarization of the muon. Notice
that integrating over $\cos \theta$, we arrive at $\Gamma (\mu^+
\to e^+ \gamma)=(m_\mu^3/4 \pi)(|A_L|^2+|A_R|^2)$.
 Thus, by measuring the total decay rate to $e^+\gamma$, we can only measure $|A_L|^2+|A_R|^2$.
However, Eq.~(\ref{averaging-spin}) shows that by studying the
angular distribution of the final particles
  with a moderate angular resolution, $|A_R|^2$ and $|A_L|^2$ can be
separately derived.  Information on $|A_L|^2/|A_R|^2$ can be
translated into information on the sources of lepton flavor
violation in the underlying theory. Studying the angular
distribution can therefore  be considered as a tool to
discriminate between different  scenarios beyond the SM
\cite{veto,review}. Moreover, in case of low statistics, studying
the angular distribution can help us to veto the background
\cite{veto}. Notice, however, that with this method only the
absolute values of $A_L$ and $A_R$ can be derived and no
information on the relative phase of $A_L$ and $A_R$  can be
extracted. Whereas the relative phase of $A_L$ and $A_R$ carry
valuable information on the sources of CP-violation in the
underlying theory. In this letter, we show that if in addition to
the angular distribution of the final particles in the LFV muon
decay, we also measure their polarization, we will be able to
extract the phase of $A_L^*A_R$. Remembering the fact that the
state-of-the-art LHC experiment will most likely not be able to
measure these phases and for measuring such phases a more advanced
collider, ILC, is proposed \cite{ILCvsLHC}, the possibility of
measuring these phases by muon decay experiments seems more
exciting. In section II, we show that by studying the angular
distribution of transversely polarized positrons and photons we
can extract the relative phase of $A_L$ and $A_R$. In section III,
we discuss the possibility of extracting the same information by
studying the angular distribution of the final positrons produced
in $\mu^+\to e^+e^-e^+$. We then compare the two methods and
discuss the advantages and disadvantages of each one. We summarize
our results in section IV.
\section{Lepton flavor violating rare decay $\mu \to e \gamma$}
Consider an anti-muon at rest [{\it i.e.,}
$P_{\mu^+}=(m_\mu,0,0,0)$] which decays into a positron and a
photon with definite spins of $\vec{s}_e$ and $\vec{s}_\gamma$,
respectively. Using the effective Lagrangian
(\ref{LFV-lagrangian}), we can calculate the $\mu \to e \gamma $
decay rate: \be \frac{d \Gamma [ \mu^+(P_{\mu^+})\to e^+(P_{e^+},
\vec{s}_{e^+}) \gamma(P_\gamma, \vec{s}_\gamma)]}{d \cos
\theta}=\frac{ m_\mu^3}{8 \pi}
\left[|\alpha_+|^2|A_L|^2(1+\mathbb{P}_\mu \cos \theta) \sin^2
\frac {\theta_s}{2}+\right.\label{no-averaging}\ee
$$ \left. |\alpha_-|^2|A_R|^2(1-\mathbb{P}_\mu \cos \theta)\cos^2
\frac{\theta_s}{2} + \mathbb{P}_\mu \textrm{Re}[\alpha_+\alpha_-^*
A_L^* A_R e^{i \phi_s} ] \sin \theta \sin \theta_s\right],
$$ where $\mathbb{P}_\mu$ is the polarization of the anti-muon,
$\theta$ is the angle between the directions of the spin of the
anti-muon and the momentum of the positron, and   $\theta_s$ is
the angle between the spin of the positron and its momentum. In
the above formula, $\phi_s$ is the azimuthal angle that the spin
of the final positron makes with the plane of spin of the muon and
the momentum of the positron (to be specific to measure $\phi_s$,
the coordinate system has been defined as follows:
$\hat{z}=\vec{p_{e^+}}/|\vec{p_{e^+}}|$ and $
\hat{y}=\vec{s}_{e^+}\times \hat{z}/|\vec{s}_{e^+}\times
\hat{z}|$). Finally, $\alpha_+$ and $\alpha_-$ give the
polarization of the final photon: $\varepsilon^\mu
=(0,\alpha_++\alpha_-,(\alpha_+-\alpha_-)i,0)/\sqrt{2}$ with
$\sqrt{|\alpha_+|^2+|\alpha_-|^2}=1$.

Summing over the spins of the final particles, we arrive at the
well-known formula shown in Eq.~(\ref{averaging-spin}) which does
not contain any information on the relative phase of $A_L$ and
$A_R$. Moreover, from Eq.~(\ref{no-averaging}) it is clear that in
order to be sensitive to the  phase of $A_LA_R^*$ the combination
$\alpha_+\alpha_-^*\sin \theta_s$ should be nonzero.
 Remember that $\alpha_-=0$ and
$\alpha_+=0$ respectively correspond to positive  and negative
helicities. On the other hand, $\sin \theta_s=0$ corresponds to
either a right-handed positron (for $\theta_s=0$) or to a
left-handed positron (for $\theta_s =\pi$). Thus, in order to
extract  the \textit{relative phase of $A_L$ and $A_R$} from
$\mu^+ \to e^+ \gamma$ we have to study the final positrons and
photons whose spins are not parallel to their momenta.

Let us now consider the CP conjugate of the same process. It is
straightforward to prove that the partial decay rate of the CP
conjugate process, $d \bar{\Gamma}/d \cos \theta$, is given by
(\ref{no-averaging}) replacing $A_L \to A_L^*$ and $ A_R \to
A_R^*$. In other words, we obtain  \be \frac{d \Gamma}{d \cos
\theta}-\frac{d \bar{\Gamma}}{d \cos
\theta}=\frac{m_\mu^3}{4\pi}\mathbb{P}_\mu
\textrm{Im}[\alpha_+\alpha_-^* e^{i \phi_s}]\textrm{Im}[A_LA_R^*]
\sin \theta \sin \theta_s. \ee As expected the difference is given
by the imaginary part of $A_L A_R^*$. Eq.~(\ref{no-averaging})
shows  that if we can run the experiment both in the muon and
anti-muon modes, we will be able to derive ${\rm Im}[A_L A_R^*]$
even without  studying  the angular distribution of the final
lepton:
$$ \int \frac{d\Gamma}{d\cos \theta}d \cos\theta-\int \frac{d \bar \Gamma}{d \cos\theta}d \cos \theta=
\frac{m_\mu^3}{8}\mathbb{P}_\mu {\rm Im} [ \alpha_+\alpha_-^* e^{i
\phi_s}] \sin \theta_s {\rm Im} [ A_L A_R^*].$$ At first sight, it
may seem that the above relation is at odds with the generalized
optical theorem \cite{weinberg} which states that the {\it total}
decay rate of a particle and an anti-particle should be equal.
Notice, however that we have not summed over the final spins so
the integrals on the left-hand side do not give the {\it total}
rate of $\mu \to e \gamma$. In fact, the above equation shows that
summing over the spin of the photon and/or the positron the
difference vanishes, as expected from the  generalized optical
theorem. The effect is maximal for linearly polarized photons
({\it i.e.,} $\alpha_-=\pm \alpha_+=1/\sqrt{2}$) and for the final
leptons polarized in the direction perpendicular to the direction
of the spin of muon and the momentum of the final lepton ({\it
i.e.,} $\theta_s=\pi/2, \phi_s=\pi/2$).

If we have only  the anti-muon mode available (or only the muon
mode available), we can still extract ${\rm Im}[A_LA_R^*]$ by
studying the angular distribution of the final leptons. Notice
that \be \int_{-1}^{-1/2}\frac{d\Gamma}{d\cos\theta}d \cos
\theta-\int_{-1/2}^{1/2} \frac{d\Gamma}{d\cos
\theta}d\cos\theta+\int_{1/2}^{1} \frac{d \Gamma}{d\cos
\theta}d\cos\theta= \frac{m_\mu^3}{8\pi}\mathbb{P}_\mu
\left(\frac{\pi}{6}-\frac{\sqrt{3}}{2}\right){\rm Re} [
\alpha_+\alpha_-^* e^{i \phi_s}A_L^* A_R] \sin \theta_s ,\ee which
shows that sensitivity to Im$[A_L^* A_R]$ is maximal again for
linearly polarized photons ({\it i.e.,} $\alpha_-=\pm
\alpha_+=1/\sqrt{2}$) and leptons polarized in the direction
perpendicular to the direction of the spin of muon and the
momentum of the final lepton ({\it i.e.,} $\theta_s=\pi/2,
\phi_s=\pi/2$).

Measuring the transverse polarization of the final lepton is
feasible. In fact, this technique has long been employed to
measure the Michel parameters \cite{transeversemichel}. Measuring
the linear polarization of photon at energies of $\sim 50$~MeV
also seems practical \cite{nim}. As recently shown in \cite{luca}
equipping the experiments with photon polarimeters can have
implications for studying the  radiative muon decay, too.

 \section{Three body decay $\mu^+ \to e^+ e^- e^+$}
  \begin{figure}[ht]
  \includegraphics[height= 5 cm,bb=5 4  470 290,clip=true]{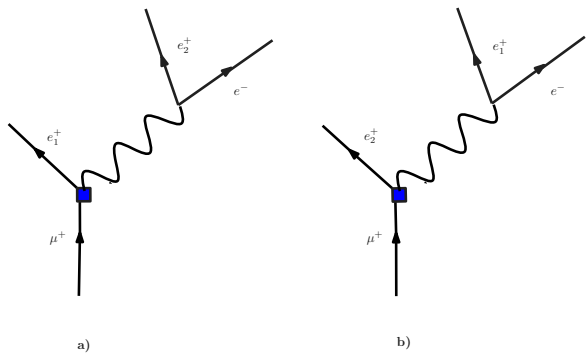}
  \caption{Penguin diagrams contributing to $\mu^+ \to e_1^+ e^+_2 e^-$. The vertices marked with boxes are
  the LFV vertices from interaction terms in Eq.~\ref{LFV-lagrangian}. } \label{diagram}
\end{figure}

 The effective
 Lagrangian in Eq.~\ref{LFV-lagrangian} can give rise to $\mu^+ \to e^+ e^- e^+$  through the penguin diagrams shown in Fig.~\ref{diagram}.
The penguin diagrams are not the only diagrams that contribute to
the decay mode $\mu^+ \to e^+ e^-e^+$ (see the appendix for more
details). However, we expect the contribution from the penguin
diagrams to be dominant because for the final lepton with energy
close $m_\mu/2$, as we will see momentarily, the photon in the
penguin diagram can go almost on shell, resulting in an
enhancement by $\ln (m_\mu /m_e)$.


 In fact, we expect about 90 \% of the three-body
$\mu^+\to e^+e^-e^+$ decays to result in a lepton with energy
$\simeq m_\mu/2$.
 In this section, we show that if we measure  the
spin of the final lepton with energy $m_\mu/2$ as well as  the
angular distributions of the final particles,  we can extract
information on the relative phase of $A_L$ and $A_R$.

 \begin{figure}[ht]\hskip -5 cm
  \includegraphics[height= 5 cm,bb=67 83 300 307,clip=true]{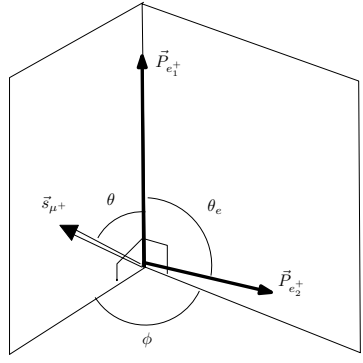}
  \caption{This figure schematically depicts the direction of the  momenta of the final particles in the LFV decay
  $\mu^+\to e^+_1 e^- e^+_2$ relative to the
  spin of the anti-muon in its rest frame.} \label{3D}
\end{figure}

Consider an anti-muon at rest with a spin at the
($\hat{x},\hat{z}$) plane which makes an angle of $\theta$ with
the $z$-axis: \be P_{\mu^+}=(m_\mu  , 0 , 0  , 0) \ \ \ \
\bar{v}_{\mu^+}=\sqrt{m_\mu} (-\sin \frac{\theta}{2}  , \cos
\frac{\theta}{2} , \sin \frac{\theta}{2} , -\cos
\frac{\theta}{2}), \label{muon-spinor}\ee where $P_{\mu^+}$ and
$v_{\mu^+}$ are respectively the four-momentum and the Dirac
spinor of the anti-muon.
 Suppose
the anti-muon decays into an electron and two  positrons with the
following momenta: \be P_{e_1^+}=(E_1,0,0,\sqrt{E_1^2-m_e^2}) \ \
\ \ P_{e_2^+}=(E_2,|\vec{P}_{e^+_2}|\sin \theta_{e}\cos
\phi,|\vec{P}_{e^+_2}|\sin \theta_e \sin \phi,|\vec{P}_{e^+_2}|
\cos \theta_e), \label{momenta}\ee where
$E_2^2=|\vec{P}_{e^+_2}|^2+m_e^2$.  The above angles are
illustrated in Fig.~\ref{3D}. The four-momentum of the final
electron is determined by the energy-momentum conservation.
  The diagrams contributing to $\mu^+\to e^+_1 e^- e^+_2$ are shown in Fig.~\ref{diagram}. The amplitude corresponding to diagram
  (\ref{diagram}-a)   can be
written as \be \bar{u}_{e} \gamma^\nu v_{e^+_2} {i g_{\nu \mu}
\over q^2} \bar{v}_{\mu^+} \sigma^{\mu \alpha} (A_L P_R +A_R P_L)
v_{e^+_1} q_\alpha,\label{ampli} \ee where $q_\alpha $ is the
four-momentum of the virtual photon in the penguin diagram:
$q\equiv P_{\mu^+}-P_{e_1^+}$. Combining Eqs.~(\ref{muon-spinor})
and (\ref{momenta}), we find $q^2=m_\mu^2+m_e^2-2E_1 m_\mu$.  As a
result, in the limit $E_1 \to m_\mu/2$, $q^2 \ll m_\mu^2$ and the
amplitude in Eq.~(\ref{ampli}) is considerably enhanced. That is
while the propagator of the virtual photon appearing in
Fig.~(\ref{diagram}-b) is given by $1/(P_{\mu^+}-P_{e^+_2})^2\sim
1/m_\mu^2$ so, in the limit $E_1 \to m_\mu/2$, the effect of
diagram (\ref{diagram}-b) can be neglected in comparison to that
of diagram (\ref{diagram}-a).  Moreover, in this limit, the
effects of the LFV terms other than the term in
Eq.~(\ref{LFV-lagrangian}) are lower at least  by a factor of
$m_e^2/m_\mu^2$ and can   be also neglected (see the appendix for
more details).
 Let us define $d \Gamma^{Max}/d
\cos \theta d \phi$ as partial decay rate of $\mu^+$ into a
positron with energy close to $m_\mu /2$ and spinor $
v_{e^+_1}=\sqrt{2E_1} ( 0 , d_e, c_e,0)^T$: \be
\label{gammamaxdef} \frac{d\Gamma^{Max}}{d \cos \theta d
\phi}=\sum_{spins} \int_{m_\mu/2 -\Delta
E}^{E_{max}}\int_{m_\mu/2-E_1}^{m_\mu/2} \frac{d \Gamma}{d E_2 d
E_1 d\cos \theta d \phi} dE_2 dE_1,\ee where $\Delta E \ll m_\mu$
and $E_{max}\simeq m_\mu/2-4 m_e^2/m_\mu$. Notice that we have
integrated and summed over the energies and spins of the pair of
$e_2^+$ and $e^-$ but not over those of $e_1^+$. It is
straightforward to show that \be \label{gammamax}\frac{d
\Gamma^{Max}}{d \cos \theta d \phi}=\frac{m_\mu^3}{768 \pi^4}
\left[ |A_L|^2|c_e|^2(1+\mathbb{P}_\mu \cos
\theta)+|A_R|^2|d_e|^2(1-\mathbb{P}_\mu \cos \theta)\right.\ee
$$\left. + \mathbb{P}_\mu \sin \theta \left( \cos (2 \phi) {\rm Re}[A_R
A_L^* d_e c_e^*] +\sin (2 \phi) {\rm Im}[A_R A_L^* d_e
c_e^*]\right) \right]\log \frac{4 m_e^2}{m_\mu \Delta E}, $$
where, as shown in Fig.~\ref{3D}, $\theta$ is the angle between
the spin of the anti-muon and the momentum of $e_1^+$ and $\phi$
is the azimuthal angle of the momentum of $e_2^+$ measured from
the plane made by the spin of $\mu^+$ and the momentum of $e^+_1$
[see Eq.~(\ref{momenta})].

After integrating over $\phi$ and $\cos \theta$ and summing over
the spin of $e^{+}_1$ ({\it i.e.,} summing over states $c_e=1,\
d_e=0$ and $d_e=1, \ c_e=0$), we will arrive at the familiar
formula in the literature ({\it e.g.,} see \cite{hisano}).
However, in this case the information on the phase of $A_R A_L^*$
will be lost. In order to extract this phase,  we have to be able
to measure the spin of $e^+_1$ as well as the direction of the
momenta of the final states relative to the spin of the anti-muon.
Let us now define the following ratio \be \label{ratio}
\mathcal{R}={\int_{-1}^{+1} d \cos \theta[\int_{0}^{2\pi} \frac{d
\Gamma^{Max}}{d \cos \theta d \phi}{\rm sgn}(\tan \phi) d \phi]
\over \int_{-1}^{+1}d \cos \theta[\int_{0}^{2\pi} \frac{d
\Gamma^{Max}}{d \cos \theta d \phi} {\rm sgn}(\tan(\phi+\pi/4)) d
\phi ] }.\ee Notice that ${\rm sgn}(\tan \phi)$ in the integral is
equal to $\pm 1$ depending on the quadrant that $\phi$ belongs to.
In principle, if the polarization of the anti-muon is large ({\it
i.e.,} $\mathbb{P}_\mu$ is about 100\%), this ratio can  be
measured in the lab. Using Eq.~(\ref{gammamax}), we can show that
$$\mathcal{R}= {{\rm Im}[ A_R A_L^* d_e c_e^*] \over {\rm Re}[A_R
A_L^* d_e c_e^*]},$$ which directly gives the phase of $A_R A_L^*$
for a transversely polarized positron, $d_e=c_e=1/\sqrt{2}$.

 Now let us compare the
advantages and disadvantages of each decay mode. In general, we
expect $${{\rm Br}(\mu^+ \to e^+ e^+e^-)\over {\rm Br}(\mu^+\to
e^+\gamma)}\simeq \frac{\alpha}{3\pi}
[\log(\frac{m_\mu^2}{m_e^2})-\frac{11}{4}]\simeq 0.0061.$$ Thus,
measurement of $\mu^+ \to e^+ e^-e^+$ will suffer from a higher
statistical uncertainty. On the other hand, to extract the
relative phase of $A_L$ and $A_R$ by studying $\mu^+ \to e^+
\gamma$ in addition to measuring the spin of $e^+$, it is
necessary to measure the spin of $\gamma$, too. Whereas in the
case of $\mu^+ \to e^+ e^-e^+$, one has to  measure only the spin
of the final positron with energy close to $m_\mu/2$.

As is well-known in the case of a three-body decay mode, we can
have CP- and T-odd  observable quantities, even if the spins of
the final particles are averaged over. However, the above
discussion shows that {\it if} the effective Lagrangian
(\ref{LFV-lagrangian}) is the {\it only} source of LFV, once we
average over the final spins, the CP- and T-odd effects will
disappear. In fact, as shown in \cite{okada,kitano}, if the
four-fermion LFV terms listed in the appendix are also present,
the CP- and T-odd effects will persist even after averaging over
the final spins. However, we generally expect these effects to be
suppressed roughly by a factor of $C_i m_\mu/[A_{L,R} \log
(4m_e^2/m_\mu \Delta E)]$ compared to the effect we have discussed
in the present paper. Notice that the two effects are sensitive to
different combinations of the CP-violating phases and can be thus
considered as complementary.

 \section{Concluding remarks}
In this paper, we have suggested a new method to derive
information on the sources of CP-violation beyond those in the SM.
The method is based on studying the rare LFV decay of polarized
muons. We have performed our analysis within a general effective
LFV Lagrangian so our results apply to any beyond  SM scenario
that violates the lepton flavor by adding new particles at
energies higher than the electroweak symmetry breaking scale ({\it
e.g.,} supersymmetry, large extra dimensions).

 We have first studied the LFV rare decay $\mu \to e
\gamma$ and shown that provided that the polarization of the final
particles are not parallel to their momentum ({\it e.g.,} if $e$
and $\gamma$ are  respectively transversely and linearly
polarized), by studying the angular distribution of $e$ and the
photon relative to the polarization of $\mu$, we can extract
information on the CP-violating phases. We have also shown that if
both muon and anti-muon modes are available, the same information
can be derived by comparing $\Gamma(\mu^+ \to
e^+_{\nshortparallel} \gamma_\nshortparallel)$ and $\Gamma(\mu^-
\to e^-_\nshortparallel \gamma_\nshortparallel)$ where the
subscript $\nshortparallel$ indicates that the spin of the
particle is not parallel to its momentum.

We have also studied the  $\mu^+ \to e^+_1 e^-e^+_2$ decay where
$e^+_1$ is defined to be the more energetic positron. The
amplitude of $\mu^+\to e^+_1 e^-e^+_2$ is severely enhanced if the
energy of $e^+_1$ is close to $m_\mu/2$. Thus, we expect the
majority of $e^+_1$ to have energies close to $m_\mu/2$. We have
focused on decays with such kinematics and proposed a new method
for extracting information on the CP-violating phases which is
based on  studying  the angular distribution of the final
particles. We have shown that with transversely polarized $e^+_1$
one can extract information on a combination of the CP-violating
phases that is impossible to achieve if   $e^+_1$ with helicity
$\pm 1$ is employed or if  the final spins are averaged over.
Notice that in this method measuring the spin of only one of the
final particles ({\it i.e.,} $e^+_1$) will be enough. We have
discussed the differences and synergies between this method and
the one discussed in \cite{okada,kitano}.

\section*{Appendix:  LFV effective Lagrangian}
In the appendix, we   discuss   possible  LFV operators that
appear by integrating out the heavy states within theories such as
the MSSM and show that the effect of Eq.~(\ref{LFV-lagrangian}) on
rare LFV  muon decays  is dominant.

  In the
literature (see {\it e.g.,} \cite{hisano}), it has been shown that
in the context of the MSSM, integrating out the heavy
supersymmetric states
 the effective LFV
Lagrangian of the electron-muon system will, in addition to
Eq.~(\ref{LFV-lagrangian}), contain   \be \varepsilon^\alpha
\bar{\mu} q^2 \gamma_\alpha (B_L P_L+B_R P_R)e +{\rm H.c.},
\label{bar}\ee where $\varepsilon^\alpha$ is the photon field, $q$
is the momentum of the photon, $P_L$ ($P_R$) is left (right)
projection matrix and $B_L$ and $B_R$ are couplings with dimension
of $[{\rm mass}]^{-2}$. This effective term will have no impact on
$\mu \to e \gamma$ simply because for on-shell photon ($q^2=0$),
this term vanishes. However, in general it can contribute to
$\mu^+\to e^+ e^- e^+$ through a penguin diagram. Notice that
unlike the case of Eq.~(\ref{ampli}), in this case the penguin
diagram does not diverge as the photon propagator goes on-shell.
Moreover, for most of the parameter space of the MSSM $B_{L,R}
m_\mu \ll A_{L,R}$ so the effect of Eq.~(\ref{bar}) is further
suppressed. As a result, for calculating $\Gamma^{Max}$ [defined
in Eq.~(\ref{gammamaxdef})] we can neglect the effect of
Eq.~(\ref{bar}).

The effective LFV effective Lagrangian will also contain the
following four-fermion terms that can in principle contribute to
$\mu^+\to e^+e^-e^+$: \ba {\cal L}=& C_1 (\bar{\mu}_R
e_L)(\bar{e}_R e_L)+C_2 (\bar{\mu}_Le_R) (\bar{e}_L e_R)\ \ \ \ \
\ \ \ \ \ \ \ \ \ \ \ \ \ \ \  \ \ \cr +& C_3(\bar{\mu}_R
\gamma^\mu e_R) (\bar{e}_R \gamma_\mu e_R)+C_4 (\bar{\mu}_L
\gamma^\mu e_L)(\bar{e}_L \gamma_\mu e_L) \ \ \ \ \ \ \ \ \ \cr +
& C_5 (\bar{\mu}_R \gamma^\mu e_R)(\bar{e}_L \gamma_\mu e_L)+C_6(
\bar{\mu}_L \gamma^\mu e_L) (\bar{e}_R\gamma_\mu e_R)+{\rm
H.c.}\ea Again, we expect that the effect of the above
four-fermion terms on $\Gamma^{max}$ [see Eq.~(\ref{gammamaxdef})]
to be negligible compared to terms in Eq.~(\ref{LFV-lagrangian}).
That is because, unlike the penguin diagrams in
Fig.~\ref{diagram}, the diagrams corresponding to the above
four-fermion interaction terms do not diverge for $E_1 \to
m_\mu/2$. Moreover, for most of the parameter space of the MSSM
the four-fermion effective couplings are small; {\it i.e.,}
$C_{i}m_\mu \ll A_{L,R}$.

One should notice that  the most general effective LFV Lagrangian,
in addition to the terms discussed above, contains extra terms.
For example, it is possible to have terms such as
$$\epsilon^{\alpha \beta \mu \nu} \bar{\mu} p_\alpha \gamma_\beta
(D_L P_L+D_RP_R)e F_{\mu \nu}+{\rm H.c.},$$ where $p_\alpha$ is
the four-momentum of the electron. However, studying the effective
LFV Lagrangian in its most general form is beyond the scope of
this letter.

\begin{acknowledgments}
I would like to thank T. Mori whose fruitful comments motivated me
to perform this analysis. I am  grateful to M. Peskin for useful
comments and encouragement. I also appreciate M. M. Sheikh-Jabbari
for careful reading of the manuscript, useful comments and
specially pointing out the discussion in the last part of the
appendix.

\end{acknowledgments}


\begin{thebibliography}{99}

\bibitem{pdg}
  W.~M.~Yao {\it et al.}  [Particle Data Group],
  J.\ Phys.\ G {\bf 33} (2006) 1.

\bibitem{MEGhomepage}
http://meg.web.psi.ch/index.html; {\it see also}
  M.~Grassi  [MEG Collaboration],
  Nucl.\ Phys.\ Proc.\ Suppl.\  {\bf 149} (2005) 369.

\bibitem{review}
  Y.~Kuno and Y.~Okada,
  Rev.\ Mod.\ Phys.\  {\bf 73}, 151 (2001)
  [arXiv:hep-ph/9909265];
 J.~L.~Feng,
  arXiv:hep-ph/0101122.


\bibitem{veto}
Y.~Kuno and Y.~Okada,
  Phys.\ Rev.\ Lett.\  {\bf 77}, 434 (1996)
  [arXiv:hep-ph/9604296].
\bibitem{ILCvsLHC}
{\it See  for example,}
  R.~M.~Godbole,
  Czech.\ J.\ Phys.\  {\bf 55} (2005) B221
  [arXiv:hep-ph/0503088];
 S.~Heinemeyer and M.~Velasco,
{\it In the Proceedings of 2005 International Linear Collider
Workshop (LCWS 2005), Stanford, California, 18-22 Mar 2005, pp
0508}
  [arXiv:hep-ph/0506267];
O.~Kittel,
  arXiv:hep-ph/0504183.
\bibitem{weinberg}
S. Weinberg, {\it The Quantum Theory of Fields}, (Cambridge
University Press), 1995, Vol 1: Section 3.6.
\bibitem{transeversemichel}
 H.~Burkard {\it et al.},
  Phys.\ Lett.\  B {\bf 160} (1985) 343.


\bibitem{nim}
 P.~F.~Bloser, S.~D.~Hunter, G.~O.~Depaola and F.~Longo,
  arXiv:astro-ph/0308331;

F. Adamyan {\it et. al,} Nucl.\ Ins.\ and Meth. in Phys. Research
A {\bf 546} (2005) 376.
\bibitem{okada}
  Y.~Okada, K.~i.~Okumura and Y.~Shimizu,
  Phys.\ Rev.\ D {\bf 61} (2000) 094001
  [arXiv:hep-ph/9906446].


\bibitem{luca}
E.~Gabrielli and L.~Trentadue,
  arXiv:hep-ph/0507191.

\bibitem{hisano}
  J.~Hisano, T.~Moroi, K.~Tobe and M.~Yamaguchi,
  Phys.\ Rev.\ D {\bf 53} (1996) 2442
  [arXiv:hep-ph/9510309].
\bibitem{kitano}
  R.~Kitano and Y.~Okada,
  Phys.\ Rev.\ D {\bf 63} (2001) 113003
  [arXiv:hep-ph/0012040].

\end{thebibliography}
\end{document}